\documentclass[aps,twocolumn,groupedaddress]{revtex4}



\usepackage{graphicx}
\usepackage{dcolumn}

\begin{document}


\title{Single Curve Collapse of the Price Impact Function for the New York Stock Exchange}




\author{Fabrizio Lillo}
\email[]{lillo@lagash.dft.unipa.it}
\affiliation{Istituto Nazionale per la Fisica della Materia, Unit\`a di 
Palermo, Viale delle Scienze, I-90128, Palermo, Italy}


\author{J. Doyne Farmer}
\email[]{jdf@santafe.edu}
\affiliation{McKinsey Professor, Santa Fe Institute, 1399 Hyde Park
Road, Santa Fe, NM 87501}

\author{Rosario N. Mantegna}
\email[]{mantegna@unipa.it}
\affiliation{Istituto Nazionale per la Fisica della Materia, Unit\`a di 
Palermo, Viale delle Scienze, I-90128, Palermo, Italy
and Dipartimento di Fisica e Tecnologie Relative, Universit\`a di Palermo,
Viale delle Scienze, I-90128, Palermo, Italy}



\begin{abstract}
We study the average price impact of a {\it single} trade executed in
the NYSE.  After appropriate averaging and
rescaling, the data for the 1000 most highly
capitalized stocks collapse onto a single function, giving average
price shift as a function of trade size.  This function increases as a
power that is the order of $1/2$ for small volumes, but then increases
more slowly for large volumes.  We obtain similar results in each year
from the period 1995 - 1998. We also find that small volume liquidity 
scales as a power of the stock capitalization.
\end{abstract}

\maketitle

Although supply and demand are perhaps the most fundamental concepts
in economics, finding any general form for their behavior has
proved to be elusive.  Here we build on earlier studies of
how trading affects prices
\cite{Hausman92,Chan95,Hasbrouck96,Farmer96,Torre97,Kempf98,chordia,Dufour2000,Rosenow01,Plerou01}.  
Our study adds to these previous efforts by using huge
amounts of data, by looking at the short term response to a single
trade, and by measuring the market activity in units of transactions rather 
than seconds, so that we can more naturally aggregate data for many
different stocks.  This allows us to find regularities in the response
of prices to new orders that have previously been hidden by the
extremely high noise levels that dominate financial prices.  We study
the 1000 largest stocks of the New York Stock Exchange, from
(1995-1998), and find that, by appropriate averaging and rescaling, it
is possible to collapse the price shift caused by a transaction onto a
single curve.  The price shift grows slowly
with transaction size, growing very roughly as the 1/2  power in
the small volume limit, and much more slowly in the large volume
limit. 
The fact that such consistent results are seen across many stocks and
for four different years suggests regularities in supply and demand.
Orders can be viewed as expressions of changes in supply and demand,
and the existence of a master price impact curve reflects the fact
that fluctuations from the supply and demand equilibrium for a large
number of financial assets, differing in economic sectors of activity
and market capitalization, are governed by the same statistical rule.

The response of prices to orders is a key property of a market.  If an
attempt to buy or sell results in a small change in price, then the
market is considered {\it liquid}; otherwise it is considered {\it
illiquid}.  One expects liquidity to depend on properties of the
asset, such as trading volume, or for stocks, the market
capitalization (the total worth of the company, i.e. the total number
of shares times their price).  The data collapse that we observe here
gives a clearer understanding of how liquidity depends on volume and
market capitalization.

The study is based on the Trades And Quotes (TAQ) database, which
contains the prices for all transactions as well as price quotations
(the best offers to buy and sell at a given price at any given time)
for the US equity markets.
We analyze data for the period 1995-1998 for the 1000 stocks with the
largest market capitalization traded in the New York Stock Exchange.
The analysis is based on roughly 113 million transactions and 173
million quotes.

Our goal is to understand how much the price changes on average in
response to an order to buy or sell of a given size.  Of course, in
each trade there is both a buyer and a seller.  Nonetheless, one often
loosely refers to a trade as a ``buy'' or a ``sell'' depending on
whether the initiator of the trade was buying or selling.  By {\it
initiator} we mean the agent who placed the more recent order. 
Buy orders tend to drive the price up, and sell
orders tend to drive it down.  It is this {\it price impact} that we
are interested in.  

Based on only transactions and quotes it is not possible to know with
certainty whether trades are initiated by buyers or sellers.  However,
we can infer this indirectly using an algorithm developed by Lee and
Ready \cite{Lee91}.  This algorithm identifies the correct sign of
trades by comparing the prices of transactions with recent quotes.
The Lee and Ready algorithm is able to classify the sign of
approximately $85\%$ of the trades of our database.  An order by a
single party may trigger transactions with multiple counterparts;
from the TAQ database we can only see transactions. To cope with this,
we lump together all transactions with the same timestamp and treat
them as a single trade.

We study the shift in the midquote price caused by the most recent
transaction.  For each transaction of volume $\omega$ occurring at
time $t$ we observe two cases: (i) When the next event is a quote
revision, we compare the next quote to the previous (prevailing)
quote, and compute the difference in the logarithm of the midquote
price.  Letting the logarithm of the midquote price be $p(t)$, we
compute the price shift $\Delta p(t_i+1) = p(t_{i+1}) - p(t_i)$, where
$t_i$ is the time of the prevailing (previous) quote and $t_{i+1}$ is
the time of the next quote following the transaction; (ii) When the
next event is a new transaction we set the price shift $\Delta p(t_i)$
to zero \cite{NoteA}.  We then investigate the average price shift as
a function of the transaction size $\omega$ measured in dollars, doing this
separately for buys and sells.

To investigate the average behavior we bin the data based on
transaction size and compute the average price shift for the data in each
bin. To put all stocks on roughly the same footing, we normalize the
transaction size by dividing by its average value for each stock in each
year.  
The results of doing this for two representative stocks are
shown in figure 1.
\begin{figure}
\includegraphics[scale=0.3, angle = -90]{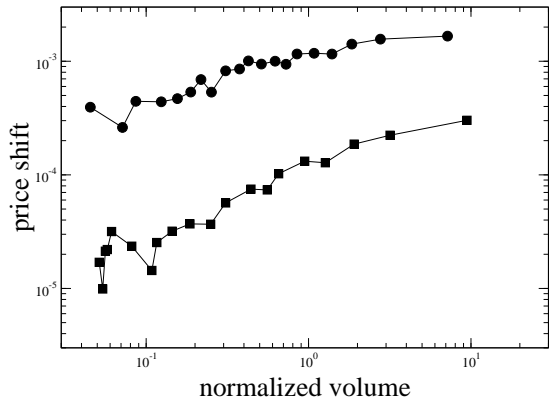}
\caption{Price shift vs. normalized transaction size for buyer 
initiated order for two representative stocks, General Electric
Co. (squares) and International Rectifier Corp. (circles) in 1995.)}
\label{twoStocks}
\end{figure}
For one of the highest cap stocks (General Electric) the average price 
impact increases roughly as $\omega^{0.6}$ throughout almost the entire
volume range .  In contrast, for a mid-cap stock 
such as International Rectifier Corp. (IRF),
for small $\omega$ ($\omega<1$) the price impact increases as
$\omega^{0.5}$, but for larger values of $\omega$ it increases roughly
as $\omega^{0.2}$. The behavior is roughly the same for both buy and
sell trades.

\begin{figure}
\includegraphics[scale=0.3, angle=-90]{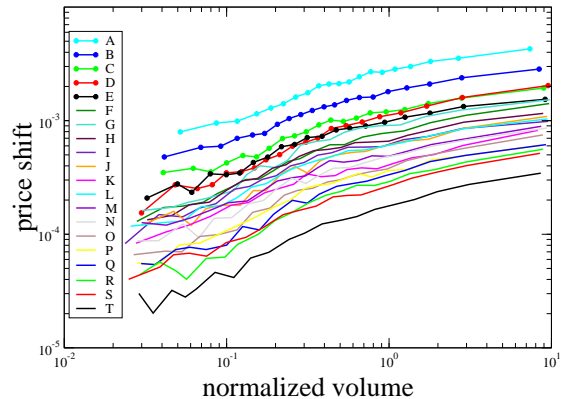}
\caption{
Price shift vs. normalized transaction size for buyer initiated trades for 20
groups of stocks sorted by market capitalization. The investigated 
year is 1995. The mean market capitalization increases from group A
to group T.}
\label{impacts}
\end{figure}

To understand more systematically how this behavior varies with market
capitalization, we group the 1000 stocks of our sample into 20 groups.
The groups are ordered by market cap, and the number of stocks in each
group is chosen to keep roughly the same number of transactions in
each group.  The groups are labeled with letters from A to T.
The group size varies from the highest market cap group (T)
with 9 stocks, to the least capitalized group (A) with 290 stocks 
\cite{Groups}.
We then bin each transaction based on size, choosing bin widths to
maintain roughly the same number of transactions in each bin 
(18,000 in 1995, 22,000 in 1996, 33,000 in 1997 and 46,000 in 1998), 
and plot the
average price impact vs. transaction size for each group.  The results
obtained for 1995 are shown in figure (\ref{impacts}).
The price impact functions naturally stratify themselves from top to bottom
in increasing order of market capitalization.  The slope of
each curve varies from roughly $0.5$ (for small transactions in higher
cap stocks) to $\approx0.2$ for larger transactions in lower cap stocks).
When we repeat this for 1996, 1997 and 1998, we see similar results, 
except that the slopes are somewhat increasingly flatter, ranging roughly 
from $\approx 0.4$ to $\approx 0.1$ in 1998.

\begin{figure}
\includegraphics[scale=0.3,angle=-90]{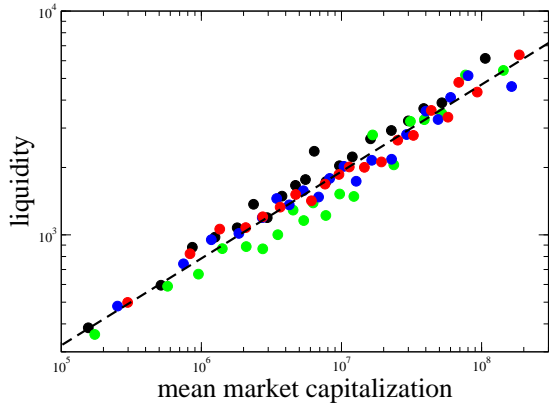}
\caption{
Liquidity $\lambda$ as a function of mean market capitalization of each
group of stocks for 1995 (black), 1996 (green), 1997 (blue) and
1998 (red). The black dashed line is the power law best fit on all points.
The best fitting exponent is 0.39.}
\label{scaling}
\end{figure}

It is clear from these results that higher market cap stocks tend to
have smaller price responses for the same normalized transaction
size. Naively, one might have expected liquidity to be proportional to
volume; the fact that the price impact for higher cap stocks is lower,
even though we are normalizing the $x$-axis by average transaction size,
says that larger cap stocks are even more liquid than one would
expect.  To gain a better understanding of this, we perform a best fit
of the impact curves for small values of the normalized transaction size
with the functional form $\Delta
p=sign(\omega)|\omega|^{\beta}/\lambda$.  In figure (\ref{scaling}) we
plot the parameter liquidity parameter $\lambda$ as a function of the
mean market capitalization of the group.
Surprisingly, for all four years the liquidity of each group increases
as roughly $C^{0.4}$, where $C$ is the average market cap of each
group (the individual values of the exponent are $0.40$, $0.42$, $0.37$,
and $0.37$ for each year, respectively.)

\begin{figure}
\includegraphics[scale=0.3,angle=-90]{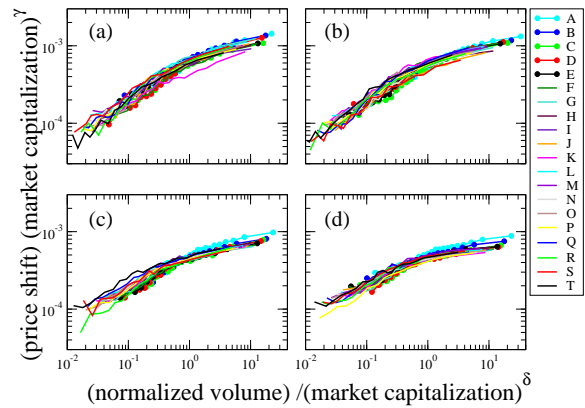}
\caption{ The price shift vs. transaction size, 
for buy orders in 1995 (a), 1996 (b), 1997 (c) and
1998 (d), renormalized as described in the text in order to make
the data collapse roughly onto a single curve.  The parameter
$\gamma=0.3$ for all years and the parameter $\delta=0.3$ for 1995,
1997, and 1998 and $\delta=0.4$ for 1996.  Results for sell orders are
very similar.}
\label{collapse} 
\end{figure}

We now make use of this apparent scaling to collapse the data of
figure (\ref{impacts}) onto a single curve.  We rescale the $x$ and
$y$ axes of each group according to the transformations
\begin{equation} 
x \rightarrow x/C^{\delta} ~~~~ y \rightarrow y~C^{\gamma} 
\end{equation} 
We then search for the values of $\delta$ and $\gamma$ that do the
best job of placing all the points on a single curve.  To do this we
divide the $x$ axis into bins, and find values that minimize the mean 
of the two dimensional variance $\epsilon
=(\sigma_y/\mu_y)^2+(\sigma_x/\mu_x)^2$, where $\sigma$ denotes the
standard deviation and $\mu$ denotes the mean, and $y$ is the
renormalized return and $x$ is the renormalized transaction size.  In
all investigated years there is a clear minimum for $\delta \approx
\gamma \approx 0.3$ (to be precise $\gamma=0.3 \pm 0.05$ for all years
and $\delta=0.3\pm 0.05$ for 1995, 1997 and 1998 whereas
$\delta=0.4\pm 0.05$ for 1996).  The resulting rescaled price impact
curves for buys in the investigated years are shown in Figure
(\ref{collapse}).

In all cases the collapse is quite good.  The resulting master
function spans three decades.  It increases slower than a power law,
and decreases more slowly in 1998 than in 1995.  The data from 1996
and 1997 show similar behavior, with the slopes decreasing steadily
from year to year.

This slow rate of increase of the price impact function shown
here is surprising.  Naive arguments predict that it should
increase at least exponentially for positive $\omega$. 
In contrast, many previous empirical
studies of price impact suggest concave behavior
\cite{Hausman92,Farmer96,Torre97,Kempf98,Rosenow01,Plerou01}. 
However, this result has not been observed universally
 \cite{Maslov01}, and none of these studies have given a clear
 indication as to functional form.  We have solved the problems 
 by focusing on the most elementary response, which is
 the price impact following a {\it single} trade, by
 analyzing a huge amount of data, aggregating across different stocks
 and by scaling the data based on market capitalization.
 
The traditional approach in economics to deriving demand curves is to
assume that agents maximize their utility under assumptions about
cognitive ability and access to information.  The standard
interpretation of our results would be that the size dependence of
price impact is due to differences in the information content of
trades.  In other words, some trades are based on more information
than others, and this is known by market participants and factored
into the price setting process.  This hypothesis suffers from the
problem that the information content of trades is difficult to assess
{\it a priori}, making the hypothesis unfalsifiable.  In
contrast, an alternative approach is to study the mechanism for making
transactions in detail, under the hypothesis that order placement and
cancellation are largely random.  This results in predictions of price
impact that are qualitatively consistent with those seen here
\cite{Daniels01,Smith02}.  If these predictions are born out 
quantitatively it will be significant in demonstrating that it is
important to model financial institutions in detail, and that for some
purposes it is may be more useful to model human behavior as random
rather than rational. 


In summary, we have demonstrated a remarkable regularity in the
immediate response of stock prices to fluctuations in supply or
demand.  In each year we are able to get a good data collapse with
similar parameters. This scaling holds for stocks with trading volumes and
market capitalizations that differ by 6 and 4 orders of magnitude
respectively. The resulting data collapse is useful because it
tells us how the liquidity of stocks varies with their market cap,
increasing as powers of market cap, in a way that is not obvious {\it
a priori}.    



\begin{thebibliography}{99}
\bibitem{NoteA} We have verified that our results are essentially unchanged
when we consider only the price shifts discussed in case (i).
\end{thebibliography}

\end{document}